\documentclass[submission,copyright,creativecommons]{eptcs}
\providecommand{\event}{ACL2 2022} 
\usepackage{breakurl}             
\usepackage{underscore}           

\usepackage{wrapfig}
\usepackage{graphicx}
\usepackage{amssymb}
\usepackage{caption}
\usepackage{amsmath}
\usepackage{amsfonts}
\usepackage{listings}
\usepackage{gensymb}

\title{A Free Group of Rotations of Rank 2}
\author{Jagadish Bapanapally\footnote{The research presented in this paper was partially supported by a grant from IOG Singapore Pte. Ltd.}\; \& Ruben Gamboa
\institute{University of Wyoming \\ Laramie, Wyoming}
\email{\{jbapanap,ruben\}@uwyo.edu}
}
\def\titlerunning{A Free Group of Rotations}
\def\authorrunning{J. Bapanapally \& R. Gamboa}
\begin{document}
\maketitle

\begin{abstract}
One of the key steps in the proof of the Banach-Tarski Theorem is the introduction of a free group of rotations. First, a free group of reduced words is generated where each element of the set is represented as an ACL2 list. Then we demonstrate that there is a one-to-one relation between the set of reduced words and a set of 3D rotations. In this paper we present a way to generate this set of reduced words and we prove group properties for this set. Then, we show a way to generate a set of 3D matrices using the set of reduced words. Finally we show a formalization of 3D rotations and prove that every element of the 3D matrices set is a rotation.
\end{abstract}

\section{Introduction}
\label{section:Introduction}
The Banach-Tarski theorem~\cite
 {weston2016banach} states that we can break the unit ball into a
 finite number of sets, then rotate the sets to form two identical
 copies of the unit ball. This seems impossible because it contradicts our
 intuition that when we partition the ball into finite sets, the total
 volume of the pieces must be the same as the volume of the original
 ball. This would be the case if all the pieces had a well-defined
 volume. The Banach-Tarski theorem is possible because the construction
 breaks the ball into non-measurable sets~\cite{jech2008axiom}, which
 means they don't have a well-defined volume. These non-measurable sets are formed by the introduction of a free group of rotations. Using properties of this free group and with the help of the Axiom of Choice, the surface of the sphere is broken down into two equivalent sets. This construction is then extended to the whole unit ball.
 
 The free group of rotations is formed by introducing a free group of reduced words and then showing one-to-one relation between the set of reduced words and the set of rotations. In section 2, we generate the set of reduced words using ACL2 lists. In section 3, we show a way to generate a set of 3D matrices using the set of reduced words. Then we show there is a one-to-one relation between the set of reduced words and the set of 3D matrices. In section 4, we formalize 3D rotations in ACL2(r) and we show every element of the 3D matrices set is a rotation thus generating a free group of rotations. Many properties of matrix algebra \cite{gamboa2003using} and modular arithmetic \cite{Bertoli2000} that are needed for the proof have already been formalized in ACL2(r). The matrix algebra that is formalized using the ACL2 two dimensional arrays contains a lot of properties that we need for the proof. For example matrix multiplication, matrix equivalence, matrix transpose, and properties like associativity of the matrix multiplication and ($m_1 \times m_2$)$^T = m_1^T \times m_2^T$ have been formalized. Also, there are properties about dimensions of the matrices. These formalized properties about matrices made us believe we can use these books and we have been proven correct as we are able to achieve the goal of generating a free group of rotations of rank 2.
 

\section{A Free Group of Reduced Words}
\label{section:reducedwords}
In this section, we introduce the free group over the letters $a$ and
$b$. This group contains all words that can be formed from $a$, $b$,
$a^{-1}$, and $b^{-1}$ such that no letter and its inverse appear
together. For example, $abba$ is a member of this free group but $abb^
{-1}a$ is not.

We use lists in ACL2(r) to represent words. A weak word is an empty list or a list that has characters $a$ or $a^{-1}$ or $b$ or $b^{-1}$. For example,
'($a$ $b$ $b^{-1}$ $a^{-1}$) is a weak word.  In the ACL2(r) source
files, we have defined the functions \texttt{wa}, \texttt{wa-inv}, \texttt{wb} and \texttt{wb-inv} which return the ACL2(r) characters $\#\backslash a$,
$\#\backslash b$, $\#\backslash c$, and $\#\backslash d$ respectively. e.g., (\texttt{wa})=$\#\backslash a$. We use the ACL2(r) characters $\#\backslash a$,
$\#\backslash b$, $\#\backslash c$, and $\#\backslash d$ to  represent
$a$, $a^{-1}$, $b$, and $b^{-1}$ respectively, but in this paper we
will simply refer to $a$, $a^{-1}$, $b$, and $b^{-1}$ to avoid
confusion. The predicate \textit{weak-wordp} recognizes elements of the
set of weak words, as shown below. Since ACL2(r) does not
have support for infinite sets, such as a set of weak words, we
represent these sets implicitly using recognizers for their elements.
\begin{lstlisting}
(defun weak-wordp (w)
  (cond ((atom w) (equal w nil))
        (t (and (or (equal (first w) (wa))
                    (equal (first w) (wa-inv))
                    (equal (first w) (wb))
                    (equal (first w) (wb-inv)))
                (weak-wordp (rest w))))))
\end{lstlisting}
A reduced word is a weak word such that character $a^{-1}$ does not
appear beside the character $a$ and character $b^{-1}$ does not appear
beside the character $b$ in the list. For instance, '($a$ $b$ $a^{-1}$) is
a reduced word and '($a$ $a^{-1}$ $b$) is not a reduced word. The
predicates \textit{a-wordp}, \textit{a-inv-wordp}, \textit
{b-wordp}, and \textit{b-inv-wordp} represent the set of reduced words
that start with characters $a$, $a^{-1}$, $b$, and $b^
{-1}$ respectively. The predicate \textit{reducedwordp}, as shown below, represents the set of all reduced words. \textit
{reducedwordp} returns true if the argument belongs to the set \textit
{a-wordp} or \textit{a-inv-wordp} or \textit{b-wordp} or \textit
{b-inv-wordp} or if it is an empty list.
\begin{lstlisting}
(defun reducedwordp (x)
  (or (a-wordp x)
      (a-inv-wordp x)
      (b-wordp x)
      (b-inv-wordp x)
      (equal x '())))
\end{lstlisting}
The function \textit{word-inverse} finds the inverse of a reduced word.
If the argument is a weak word, \textit{word-inverse} flips each
character in the list to its inverse and then reverses the list,
e.g., \textit{word-inverse}('($a$ $a^{-1}$ $b^{-1}$)) =  '($b$ $a$ $a^
{-1}$). Below are the definitions of the flip function and
the inverse function. 
\begin{lstlisting}
;; Definition of the flip function
(defun word-flip (x)
  (cond ((atom x) nil)
        ((equal (car x) (wa)) (cons (wa-inv) (word-flip (cdr x))))
        ((equal (car x) (wa-inv)) (cons (wa) (word-flip (cdr x))))
        ((equal (car x) (wb)) (cons (wb-inv) (word-flip (cdr x))))
        ((equal (car x) (wb-inv)) (cons (wb) (word-flip (cdr x))))))

;; Definition of the Inverse operation
(defun word-inverse (x)
  (rev (word-flip x)))
\end{lstlisting}
The group operation \textit{compose} takes two arguments. If the
arguments are weak words, then the \textit{compose} function first
appends the two lists and then ``fixes'' the result by deleting any
letter and its inverse that appear beside each other. Thus, the final
result of \textit{compose} is always a reduced word. E.g., $\mathit{compose}$('($a$ $b$ $b$), '($b^{-1}$)) = '($a$ $b$). Below are the definitions of the fixing function and the group
operation \textit{compose}. 
\begin{lstlisting}
;; Definition of the fixing function
(defun word-fix (w)
  (if (atom w)
      nil
    (let ((fixword (word-fix (cdr w))))
      (let ((w (cons (car w) fixword)))
        (cond ((equal fixword nil)
               (list (car w)))
              ((equal (car (cdr w)) (wa))
               (if (equal (car w) (wa-inv))
                   (cdr (cdr w))
                   w))
              ((equal (car (cdr w)) (wa-inv))
               (if (equal (car w) (wa))
                   (cdr (cdr w))
                   w))
              ((equal (car (cdr w)) (wb))
               (if (equal (car w) (wb-inv))
                   (cdr (cdr w))
                   w))
              ((equal (car (cdr w)) (wb-inv))
               (if (equal (car w) (wb))
                   (cdr (cdr w))
                   w)))))))

(defun compose (x y)
  (word-fix (append x y)))
\end{lstlisting}

If we denote the set of reduced words by $W(a,b)$, the set of reduced words
starting with character $a$ by $W(a)$, and similarly for $W(a^
{-1})$, $W(b)$, and $W(b^{-1})$, then $$W{(a,b)} = 
\text{'}() \; \cup\; W(a) \; \cup \; W(a^{-1}) \; \cup W
(b) \; \cup \; W(b^{-1})$$
Considering the empty list as the identity element, we show below the group properties of the set of reduced words.

\subsection{Closure Property}
If  $x$ and $y$ are reduced words, then (append $x$ $y$) is a weak
word as shown below by the lemma \textit{closure-lemma}. If $x$ is a weak word, then \textit{word-fix}($x$) returns a reduced
word as shown below by the \textit{weak-wordp-equivalent} lemma. So, \textit{compose}($x,y$) = \textit{word-fix}(append $x$
$y$) is a reduced word. This establishes that \textit{compose} is
closed over the set of  reduced words as shown below by the lemma \textit{closure-prop}.
\begin{lstlisting}
(defthmd closure-lemma
  (implies (and (reducedwordp x)
                (reducedwordp y))
           (weak-wordp (append x y))))
  
(defthmd weak-wordp-equivalent
  (implies (weak-wordp x)
           (reducedwordp (word-fix x))))

(defthmd closure-prop
  (implies (and (reducedwordp x)
                (reducedwordp y))
           (reducedwordp (compose x y))))
\end{lstlisting}
\subsection{Associative Property}
By the definition, \textit{word-fix} "fixes" a weak word recursively starting from the tail of the list; i.e if $x$, $y$, $z$ are weak words, then (\textit{word-fix} (append $x$ (\textit{word-fix} (append $y$ $z$)))) is equal to (\textit{word-fix} (append $x$ $y$ $z$)) as shown below by the lemma \textit{compose-assoc-lemma1}. Another key lemma required to prove that the set $W(a,b)$ satisfies the associative property is that if $x$ is a reduced word, then \textit
{word-fix}(rev($x$)) = (rev(\textit{word-fix}($x$))), which we proved by induction on $x$.
\begin{lstlisting}
(defthm compose-assoc-lemma1
  (implies (and (weak-wordp x)
                (weak-wordp y)
                (weak-wordp z))
           (equal (word-fix (append x (word-fix (append y z)))) 
                  (word-fix (append x y z))))
  :hints ...)
\end{lstlisting}
The other two lemmas required to prove the associative property which are already proved in ACL2, are: if $x$ and $y$ are lists, then rev(rev $x$) = $x$ and rev(append $x$ $y$) = (append (rev $y$) (rev $x$)). Using these lemmas, below is the derivation of the associative property of the \textit{compose} function with respect to the set of reduced words. If $x$, $y$, $z$ are reduced words, then
\begin{align*} 
\text{(\textit{compose} (\textit{compose} $x$ $y$) $z$)} &=  \text{(\textit{word-fix} (append (\textit{word-fix} (append $x$ $y$)) $z$))} \\ 
 &= \text{(rev (rev (\textit{word-fix} (append (\textit{word-fix} (append $x$ $y$)) $z$))))} \\
 &= \text{(rev (\textit{word-fix} (rev (append (\textit{word-fix} (append $x$ $y$)) $z$))))} \\
 &= \text{(rev (\textit{word-fix} (append (rev $z$) (rev (\textit{word-fix} (append $x$ $y$))))))} \\
 &= \text{(rev (\textit{word-fix} (append (rev $z$) (\textit{word-fix} (rev (append $x$ $y$))))))} \\
 &= \text{(rev (\textit{word-fix} (append (rev $z$) (\textit{word-fix} (append (rev $y$) (rev $x$))))))} \\
 &= \text{(rev (\textit{word-fix} (append (rev $z$) (rev $y$) (rev $x$))))} \\
 &= \text{(\textit{word-fix} (rev (append (rev $z$) (rev $y$) (rev $x$))))} \\
 &= \text{(\textit{word-fix} (append $x$ $y$ $z$))} \\
 &= \text{(\textit{word-fix} (append $x$ (\textit{word-fix} (append $y$ $z$))))} \\
 &= \text{(\textit{compose} $x$ (\textit{compose} $y$ $z$))}
\end{align*}

\subsection{Inverse Property}
By induction on $x$, first we show if $x$ is a reduced word, then (rev $x$) and (\textit{word-flip} $x$) are reduced words and thus (\textit{word-inverse} $x$) is a reduced word. Now since (\textit{word-inverse} $x$) is a reduced word, using the associative property and by induction on $x$, (\textit{compose} $x$ (\textit{word-inverse} $x$)) results in an empty list as shown below by the \textit{reduced-inverse} lemma below. This proves that right inverse of the reduced word $x$ is (\textit{word-inverse} $x$). To prove the left inverse of $x$ is also equal to (\textit{word-inverse} $x$), we can use the \textit{reduced-inverse} lemma. In the \textit{reduced-inverse} lemma in place of $x$ if we have (\textit{word-inverse} $x$) and if (\textit{word-inverse} (\textit{word-inverse} $x$)) is equal to $x$, then (\textit{word-inverse} $x$) becomes the left inverse of $x$.  We have proved (\textit{word-inverse} (\textit{word-inverse} $x$)) is equal to $x$ by functionally instantiating the \textit{equal-by-nths} \cite{Equal-by-nths} lemma. We have functionally instantiated the \textit{equal-by-nths} lemma with the hypothesis being $x$ a weak word, left hand side of the equivalence being (\textit{word-inverse} (\textit{word-inverse} $x$)) and the right hand side of the equivalence being just $x$. To finish the proof, we needed proofs that both the lists (\textit{word-inverse} (\textit{word-inverse} $x$)) and $x$ have same characters at any specified index and they both have the same length. Thus we have proved that for every element $x$ in the reduced word set there exists an inverse of $x$ which is equal to (\textit{word-inverse} x).
\begin{lstlisting}
(defthmd reduced-inverse
  (implies (reducedwordp x)
           (equal (compose x (word-inverse x)) '()))
  :hints ...)
\end{lstlisting}

\section{A Free Group of 3D Matrices}
\label{section:free-group}
Matrices in ACL2 are represented with the data structure \textit{array2p}.
We define a predicate \textit{r3-matrixp} that recognizes the set of 
3D matrices: \textit{r3-matrixp} returns true if the argument is of 
type \textit{array2p}, if its dimensions are $3\times3$, and if each 
element of the matrix is a real number.

We now define the four matrices $A^+$, $A^-$, $B^+$, and $B^-$ as
$$
A^{\pm} = \begin{bmatrix} 
1 & 0 & 0 \\
0 & \frac{1}{3} & \mp\frac{2\sqrt{2}}{3} \\
0 & \pm\frac{2\sqrt{2}}{3} & \frac{1}{3}
\end{bmatrix} \; \; \;B^{\pm} = \begin{bmatrix} 
\frac{1}{3} & \mp\frac{2\sqrt{2}}{3} & 0 \\ 
\pm\frac{2\sqrt{2}}{3} & \frac{1}{3} & 0 \\ 
0 & 0 & 1  
\end{bmatrix}
$$
and we associate these matrices with the letters $a$, $a^{-1}$, $b$, and
$b^{-1}$ from the free group respectively. Moreover, we associate a
list $(x_1, x_2, \dots, x_n) \in W(a,b)$ with the matrix $X_1 \times
X_2 \times \dots \times X_n$, where $\times$ denotes matrix
multiplication, and $X_i$ is the matrix associated with letter $x_i$.
The recursive function \textit{rotation} performs this mapping from
words in the free group to 3D matrices. If we denote the resulting set as $R
(a,b)$, then $R(a,b) = \{\mathit{rotation}(w) \mid w \in W
(a,b)\}$. By induction, it is easy to verify that every element of the set
$R(a,b)$ belongs to \textit{r3-matrixp}. 

To show the set $R(a,b)$ is a free group homomorphic to $W(a,b)$, 
we show that if $w \in W(a,b)$ and $w$ is not the empty list, 
then $\mathit{rotation}(w)$ is not equal to $I$, the identity
matrix. Equivalently, we show that $\mathit({rotation}(w))(0,1,0) 
\ne (0,1,0)$, unless $w$ is the empty list.

To do this, suppose that $w\in R(a,b)$, and consider 
the rotation $R(w)$. In particular, suppose that $R(w)$ transposes
the point $(0,1,0)$ to $(x',y',z')$. Define $(x,y,z)$ as
$$(x,y,z) = 3^n\left(\frac{x'}{\sqrt{2}},y',\frac{z'}{\sqrt{2}}\right)$$
where $n=|w|$. Using induction, we show $x$, $y$, and $z$
are integers.

So now suppose that $\mathit(rotation(w))(0,1,0)=(0,1,0)$ for
some non-empty word $w$. It
follows that $(x,y,z) = (0, 3^n, 0)$, where $n=|w|>0$, thus
$x\equiv y \equiv z \equiv 0 \pmod 3$. But this cannot be the
case. If $|w|=1$, then $\mathit rotation(w)$ is one of $A^\pm$ or
$B^\pm$, and considering each of the four cases by brute force,
it is clear that $(x,y,z) \not\equiv (0, 0, 0) \pmod 3$. Using induction, there
are 16 cases to consider, but in all of these cases we can again
conclude that $(x,y,z) \not\equiv (0, 0, 0) \pmod 3$. This shows
that if $|w|>0$, then $rotation(w)$ is not the identity matrix.

Here we want to mention two key lemmas needed to
prove the one-to-one relation between the set of 3D matrices and the set of reduced words. First, if $w_1, w_2\in W(a,b)$, then by 
the definition of \textit{rotation} and \textit{compose}, 
$\mathit{rotation}(w_1) \times \mathit{rotation}(w_2) = 
\mathit{rotation}(\mathit{compose}(w_1, w_2))$. 
Second, if $r\in R(a,b)$, then $\exists w\in W(a,b)$ such that
$r=\mathit{rotation}(w)$, and by the previous lemma,
$r^{-1}=\mathit{rotation}(w^{-1})$. Moreover, since
$w^{-1} \in W(a,b)$, $r^{-1} \in R(a,b)$. Now, if $w_1, w_2\in W(a,b)$ and $w_1 \neq w_2$ and $r_1=\mathit{rotation}(w_1)$ and $r_2=\mathit{rotation}(w_2)$, then using the proof that, if $|w|>0$, then $rotation(w)\neq I$, $r_1 \times r_2^{-1} \neq I$, which implies $r_1 \neq r_2$. This proves there is a 
one-to-one relation between the set $R(a,b)$ and the set $W(a,b)$. 
So defining
$R(a) = \{\textit{rotation}(w) \mid  w \in W(a)\}$,
$R(a^{-1}) = \{\textit{rotation}(w) \mid w \in W(a^{-1})\}$,
$R(b) = \{\textit{rotation}(w) \mid w \in W(b)\}$,
and $R(b^{-1}) = \{\textit{rotation}(w) \mid w \in W(b^{-1})\}$, then
the set of rotations $R(a,b)$ can be partitioned as
$$R(a,b) = I \; \cup \; R(a) \; \cup \; R(a^{-1}) \; \cup R(b) \; \cup R(b^{-1}).$$

\section{A Free Group of Rotations of Rank 2}
In this section we formalize 3D rotations and prove every element of the 3D matrices set is a rotation. As discussed previously, the matrix transpose operation 
(\textit{m-trans}) was formalized in prior 
work~\cite{gamboa2003using}, and as part of that, it was shown that
$(A \times B)^T = B^T \times A^T$.

We extended that formalization by introducing the function 
\textit{r3-m-determinant} that computes the determinant of a
matrix, the function \textit{r3-m-inverse} that computes 
the inverse of a 3D matrix (when possible). Using these functions,
we defined the predicate \textit{r3-rotationp} that recognizes
rotations in $\mathbb{R}^3$.
A matrix $M$ is a rotation matrix if it
satisfies these conditions~\cite{banachmadeline}:
\begin{itemize}
    \item $M$ is a 3D matrix,
    \item $M^{-1} = M^T$, and
    \item $\det(M) = 1$.
\end{itemize}

Another important detail is that every element of $R(a,b)$ must
be a rotation of $\mathbb{R}^3$. Given the correspondence between
$R(a,b)$ and $W(a,b)$ established in section \ref{section:free-group},
what we need to show is that for any $w\in W(a,b)$,
$\mathit{rotation}(w)$ satisfies the axioms of a rotation. This was done using induction on the list
$w$. It is easy to verify that the base cases are rotations; i.e.,
$I$, $A^+$, $A^{-}$, $B$ and $B^{-}$ are all rotation matrices. 
For the induction to go through, the lemma we need to prove 
$\mathit{rotation}(xw)$ is a rotation in $\mathbb{R}^3$ given that
$\mathit{rotation}(w)$ is a rotation, is that the product of
two rotation matrices $M_1$ and $M_2$ is also a rotation matrix. Below is the the proof of this lemma, and some other lemmas from matrix algebra that we proved in ACL2(r).

\begin{itemize}
\item $\text{\textit{r3-matrixp}}(m_1) \wedge \text{\textit{r3-matrixp}}(m_2)
    \implies \text{\textit{r3-matrixp}}(m_1\times m_2)$

\item    
    $\text{\textit{r3-matrixp}}(m_1) \wedge \text{\textit{r3-matrixp}}(m_2)
    \implies \det(m_1 \times m_2)=\det(m_1) \cdot det(m_2)$
    
\item
$\text{\textit{r3-matrixp}}(m)
    \implies m\times I = I \times m = m$
    
\item
$\text{\textit{r3-matrixp}}(m) \wedge \det(m)\ne0
    \implies m\times m^{-1} = m^{-1}\times m = I$
    
\item
$\text{\textit{r3-matrixp}}(m_1) \wedge \det(m_1)\ne0
  \wedge \text{\textit{r3-matrixp}}(m_2) \wedge \det(m_2)\ne0
    \\\quad\quad\implies (m_1 \times m_2)^{-1} = m_2^{-1} \times m_1^{-1}$
    
\item
$\text{\textit{r3-rotationp}}(m_1) \wedge \text{\textit{r3-rotationp}}(m_2) 
    \implies \text{\textit{r3-rotationp}}(m_1\times m_2)$
    
\item
$\text{\textit{r3-rotationp}}(m) \implies \text{\textit{r3-rotationp}}(m^{-1})$    
    
\item Rotations preserve distances ~\cite{Equal-by-nths1}. Let $p_1 = (x_1, y_1, z_1)$ and $R$ be a 
rotation matrix, and consider $p_2 = Rp_1 = (x_2, y_2, z_2)$.
Using the previous lemmas,
\begin{align*}
x_1^2 + y_1^2 + z_1^2 &= p_1^T \times p_1 \\
&= p_1^T \times (I \times p_1) \\
&= p_1^T \times ((R^{-1} \times R) \times p_1) \\
&= p_1^T \times ((R^{T} \times R) \times p_1) \\
&= (p_1^T \times R^{T}) \times (R \times p_1) \\
&= (R \times p_1)^{T} \times (R \times p_1) \\
&= p_2^T \times p_2 \\
&= x_2^2 + y_2^2 + z_2^2.
\end{align*}
\end{itemize}

\section{Conclusion}
In this paper we presented a way to generate the free group of reduced words using ACL2 lists. Using this set we have generated a free group of 3D matrices. Then we have shown a formalization of 3D rotations in ACL2(r) and we proved that every element of the 3D matrices set is a 3D rotation. When we apply these rotations on $S^2$, then with the help of the Axiom of Choice we can form two copies of $S^2$ minus the set of the poles of the rotations. This is called the Hausdorff's Paradox which is the next step in the proof of the Banach-Tarski theorem. We are currently working to formalize the Hausdorff's paradox, and then we will prove the Banach-Tarski theorem.  
\nocite{*}
\bibliographystyle{eptcs}
\bibliography{generic}
\end{document}